\documentclass[prd,amsmath,amssymb,longbibliography,superscriptaddress,twocolumn,nofootinbib,10pt]{revtex4}

\usepackage{dcolumn}
\usepackage{bm}
\usepackage{amssymb}
\usepackage{latexsym}
\usepackage{booktabs}
\usepackage{amsmath}
\usepackage{multirow}
\usepackage{graphicx} 
\usepackage{epstopdf}

\usepackage[colorlinks=true, linkcolor=blue, citecolor=blue]{hyperref}
\newcommand{\beq}[1]{\begin{equation}\label{#1}}
 \newcommand{\eeq}{\end{equation}}
 \newcommand{\bea}{\begin{eqnarray}}
 \newcommand{\eea}{\end{eqnarray}}

\def\aj{Astron. J.}
\def\aap{Astron. Astrophys.}
\def\mnras{Mon. Not. Roy. Astron. Soc.}
\def\apjl{Astrophys. J. Lett.}
\def\jcap{JCAP}
\def\apjs{Astrophys. J., Suppl. Ser.}

\begin{document}

\title{Constraining on the non-standard cosmological models combining the observations of high-redshift quasars and BAO}

\author{Ziqiang Liu}
\affiliation{School of Physics and Optoelectronic, Yangtze University, Jingzhou 434023, China}

\author{Tonghua Liu}
\email{liutongh@yangtzeu.edu.cn}
\affiliation{School of Physics and Optoelectronic, Yangtze University, Jingzhou 434023, China}

\author{Xinyi Zhong}
\affiliation{School of Physics and Optoelectronic, Yangtze University, Jingzhou 434023, China}

\author{Yifei Xu}
\affiliation{School of Physics and Optoelectronic, Yangtze University, Jingzhou 434023, China}

\author{Xiaogang Zheng}\email{xiaogang.zheng@whpu.edu.cn}
\affiliation{School of Electrical and Electronic Engineering,
	Wuhan Polytechnic University, Wuhan 430023, China}

\begin{abstract}
In this work, we studied four types of cosmological models with different mechanisms driving the accelerated expansion of the universe, include Braneworld models, Chaplygin Gas models, Emergent Dark Energy models, and cosmological torsion models. Considering that the dynamics of these models at low redshifts are very similar and difficult to distinguish, we used the latest and largest UV and X-ray measurements of quasars (QSOs) observations covering the range of redshift $0.009<z<7.5$. However, the high intrinsic dispersion of this sample and the degeneracy between cosmological model parameters, we added 2D-BAO and 3D-BAO datasets to help us constrain the parameters of these cosmological models. Our results suggest that standard cold dark matter scenario may not be the best cosmological model preferred by the high-redshift observations. The Generalized Chaplygin Gas (GCG)  and cosmological constant plus torsion (named Case II) models perform best by Akaike Information Criterion (AIC), but the $\Lambda$CDM is the best cosmological model preferred by Bayesian Information Criterion (BIC). Our work also supports that the Phenomenologically Emergent Dark Energy  and cosmological torsion models may alleviate the Hubble tension, the reported value of the Hubble constant obtained from QSO+BAO datasets combination lies between Planck 2018 observations and local measurements from the SH0ES collaboration, while other cosmological models all support that the Hubble constant tends to be closer to recent Planck 2018 results, but these model are penalized by information criterion.
\end{abstract}

\maketitle

\section{Introduction}           

\label{sect:intro}

In the end of the 20th century, two international supernova
research groups, High Redshift Supernova Team (HST) and
Supernova Cosmology Project (SCP), discovered the accelerating expansion of the distant supernova type Ia (SN Ia) \citep{Perlmutter1999,Riess1998}. In subsequent years, other cosmological observations such as cosmic microwave background radiation (CMB) \citep{Spergel2003}, Baryon Acoustic
Oscillations (BAO) \citep{Eisenstein2005,Percival2007}, largescale structure \citep{Tegmark2004} had confirmed this phenomenon.
Many cosmological models have been proposed in order to explain the accelerating expansion of the universe, such as Lambda Cold Dark Matter ($\Lambda$CDM) model, which is consistent with many cosmological observations
\citep{Alam2017,Allen2008,Cao2012,Farooq2017,Scolnic2018}, is called  standard cosmological model. However,  this model still faces many theoretical problems such as the well-known fine-tuning and coincidence problems \citep{Dalal2001,Weinberg1989}. With the rapid development of observation technology, the astronomical data obtained are becoming more and more accurate, more problems that $\Lambda$CDM model can not solve are gradually emerging, such as Hubble tension \citep{Freedman2017} and the cosmic curvature problems
\citep{Di Valentino2020,Di Valentino2021}. Therefore, there has been increasing attention to other cosmological models beyond $\Lambda$CDM, including dynamic dark energy models \citep{Boisseau2000,Kamenshchik2001}, interacting dark
energy model \citep{Amendola2000,Caldera-Cabral2009} and scalar
field theories \citep{Caldwell2005,Chen2011,Zlatev1999}. Moreover, a Phenomenologically Emergent Dark Energy (PEDE) presented by Li et al. has obtained a great
attention \citep{Li2019}, which demonstrated its potential in addressing the Hubble constant problem. Furthermore, modifications to General Relativity (GR) theory  \citep{Tsujikawa2010} could also provide a useful way to deal with the cosmological constant problem and explain the late-time acceleration of the universe without dark energy. For example, braneworld  Dvali-Gabadadze-Porrati (DGP) \citep{Dvali2000,Sollerman2009} model and modified polytropic Cardassian (MPC) model \citep{Magana2015,Wang2003}, which suppose our universe is embedded in a higher dimension of space-time to modify gravity. Similarly, the field equation can be allowed to extend the general theory of relativity to be greater than the second order, like $f(R)$ gravity \citep{Chiba2003,Sotiriou2010}. In this paper, we will study four kinds of typical cosmological models that work in the framework of Friedman-Lema\^{\i}tre-Robertson-Walker metric, including braneworld models: DGP  and a phenomenological extension DGP model, $\alpha$ dark energy ($\alpha$DE) \citep{Dvali2003}, Generalized Chaplygin Gas (GCG) model \citep{Bento2002}, a kind of dynamical dark energy model, in which the dark energy density decreases with time, and New Generalized Chaplygin Gas (NGCG) model \citep{Zhang2006}, Emergent Dark Energy models, here we consider two models: Phenomenologically
Emergent Dark Energy (PEDE) \citep{Li2019} and Generalised Emergent Dark Energy (GEDE) \citep{Li2020}, cosmological models with torsion (vanishing dark energy term called Case I and containing the dark energy term called Case II) \citep{Pereira2022}. However, given the number and precision of previous observational datasets, including the Hubble parameter \citep{Wu2007} and SN Ia \citep{Nesseris2005,Scolnic2018,Suzuki2012}, the vast majority of cosmological models are not excluded by the data at low redshifts. This is because their dynamics in the low redshift range are so similar that they are difficult to distinguish. Nevertheless, their dynamics are quite different at high redshifts. Therefore, there is an urgent need for high-redshift cosmological probes, such as quasars, to perform observational tests and place constraints on the models.

Quasars (QSOs) are among the brightest sources of stars in the universe. However, quasars do not have a normalizable luminosity, and their spectral properties and photometric correlation are weak, with large dispersion and bias. Nevertheless, if the quasar sample is large enough, it can be useful for cosmological measurements in principle. Thanks to the advancements and progress made in modern science and technology, the number of observed quasars has increased by an order of magnitude. For example, quasar samples of UV and X-ray radiation obtained from multiple optical band ROSAT, Chandra, Sloan Digital Sky Survey (SDSS), and XMM-Newton Serendipitous Source Catalogue (XMM-NSSC) sky surveys \citep{Just2007,Lusso2010,Steffen2006}, and samples of quasars of milliarc second angular size in the radio band obtained from the very-long baseline interferometry (VLBI) observations \citep{Gurvits1994,Kellermann1993}.   Quasars have also been utilized as standard candles whose standardization relies on the linear relationship between the logarithms of their ultraviolet (UV) and X-ray luminosities \citep{Geng2020,Khadka2020,Liu2020a,Liu2020b,2020ApJ...901..129L,Lusso2019,
Lusso2020,Risaliti2017,Risaliti2019,Salvestrini2019}.
In general, the advantage of quasar measurements compared to other conventional cosmological probes lies in their larger redshift range. This could be beneficial for exploring the behaviour of non-standard cosmological models at high redshift and could provide an important addition to other astrophysical observations, demonstrating the potential of quasars as an additional cosmological probe \citep{Zheng2021}.

In the previous work \cite{2021MNRAS.505.2111L}, the authors indicated that the results show that using QSO data alone are not able to provide tight constraints on model parameters, which is mainly related to the large dispersion in QSO data. In addition, they demonstrated that the use of Baryon Acoustic Oscillation (BAO) data is complementary to the QSO distance measurements. Thus, in order to get more stringent constraints on cosmological parameters, we also consider measurements of transversal BAO scale (namely 2-D BAO) \cite{Nunes2020} which come from  public data releases (DR) of the SDSS \cite{2016PhRvD..93b3530C,Alcaniz,2020APh...11902432C,2018JCAP...04..064D,2020MNRAS.492.4469D},
and BAO measurements from the post-reconstructed power spectra of the BOSS DR12 data (namely 3-D BAO) \cite{Alam2021} which come from  Main Galaxy Sample (MGS), the two BOSS galaxy samples, eBOSS luminous red galaxies (LRG), and eBOSS emission line galaxies (ELG) \cite{Alam2017,2015MNRAS.449..848H,2015MNRAS.449..835R,2021MNRAS.500..736B,2020MNRAS.498.2492G,2020MNRAS.499.5527T,
2021MNRAS.501.5616D,2021MNRAS.500.1201H,2020MNRAS.499..210N}.  These datasets have been extensively used for cosmological researches and applications in the literature \cite{2023arXiv230311271S,2023ApJ...951...63D,
2023PhRvD.108b3504C,2023MNRAS.521.5013S,2022A&A...668A.135S,2023MNRAS.521.3909B,2022MNRAS.513.5686C}. For instance,  the work  \cite{2023PhRvD.107j3521C} also used 3-D BAO data combined with other lower-redshift data (including  QSO angular size, Pantheon+ SNe, Hubble parameter datasets etc.) to constrain non-standard cosmological models, and showed the complementary role of BAO data.  However, there are some differences between the two sets of BAO data on the model constraining capabilities exhibited in different cosmological models. Recently, the work \cite{2023PhRvD.107j3531B} found that 2D-BAO measurements generate a little high $H_0$ compared to 3D-BAO measurements in the interacting dark energy model.

Inspired by the above, we combine QSO data of UV and the X-ray flux measurements and 2D and 3D BAO datasets to  constrain the cosmological models mentioned above, respectively.
The main purpose is to test the high-redshift correspondence between quasar data and the standard  $\Lambda$CDM by examining the performance of these cosmological models at high redshift. We concentrate on the nonlinear relationship between the UV and X-ray luminosities of quasars serving as standard candles across a redshift range spanning $0.009 < z < 7.5$. In Section 2, we introduce all the cosmological models.  We give  the methodology  and observation datasets including QSO plus two BAO datasets in Section 3. The  results of constrained cosmological parameter and model comparison criterion are given in Section 4. Finally, Section 5 summarizes our conclusion.

\section{Cosmology models}
The foundation of modern cosmology rests on the fundamental principles of cosmology, which state that our universe is homogeneous and isotropic on large scales. The space-time of our universe can be accurately described by the Friedmann-Lematre-Robertson-Walker (FLRW) metric:
\begin{equation}\label{eq1}
ds^2=dt^2-\frac{a(t)^2}{1-Kr^2}dr^2-a(t)^2r^2d\Omega^2,
\end{equation}
where $a(t)$ is the scale factor, and $K$ is dimensionless curvature taking one of three values $\{-1, 0, 1\}$ corresponding to a close, flat and open universe, respectively. The cosmic curvature parameter $\rm\Omega_K$ is related to $K$ and the Hubble constant $H_0$, as $\rm{\Omega_K}$ $=-c^2K/a^2_0H_0^2$.
The comoving distance $D_C$ is related to the evolution of the scale factor
\begin{equation}\label{eq2}
D_C=c\int^z_0\frac{dz'}{H(z')},
\end{equation}
where the redshift $z$ replaces the scale factor via relation $a=1/(1+z)$, the $c$ represents the speed of light, and the $H(z')$ denotes the Hubble parameter at redshift $z'$. However, we cannot actually obtain the comoving distance from observations. In fact, the cosmological distances we observe are luminosity or angular diameter distances. The dimensionless curvature $k=-1, 0 ,+1$ corresponds to open, flat and closed universes, respectively.
With such a metric, the luminosity  distance  $D_L$ and angular diameter distance $D_A$ can be expressed as \citep{2008cosm.book.....W}
\begin{equation}\label{eq2}
\frac{D_L}{1+z}={D_A}({1+z})= \left\lbrace \begin{array}{lll}
\frac{D_H}{\sqrt{|\Omega_{\rm k}|}}S_k \left(\sqrt{|\Omega_{\rm k}|}\int_{0}^{z}\frac{dz'}{E(z')}\right),\\
{D_H}S_k \left(\int_{0}^{z}\frac{dz'}{E(z')}\right), \\
\frac{D_H}{\sqrt{|\Omega_{\rm k}|}}S_k \left(\sqrt{|\Omega_{\rm k}|}\int_{0}^{z}\frac{dz'}{E(z')}\right),\\
\end{array} \right.
\end{equation}
where $D_H=c/H_0$ is known as the Hubble distance, the dimensionless Hubble parameter $E(z)$ is defined as $H(z)/H_0$. The curvature parameter $\Omega_k$ is related to $k$ as
$\Omega_k=-c^2 k/(a_0H_0)^2$, where $c$ is the speed of light. For convenience, we denoted $S_k(x)=\sin(x), x, \sinh(x)$ for $\Omega_k<0, =0, >0$.

The different cosmological theoretical models correspond to different $H(z)$  functions. For the sake of simplicity, we present the $H(z)$  functions of four cosmological models based on different theories that we aim to investigate in this paper. These models include Braneworld models (DGP and $\alpha$DE), Chaplygin Gas models (GCG and NGCG), Emergent Dark Energy models (PEDE and GEDE), and cosmological models with torsion (vanishing dark energy term called Case I and containing the dark energy term called Case II). Given that there is a strong degeneracy between dark energy and the cosmic curvature, and the latest Planck observations support the flatness of our universe with extremely high precision, we consider a flat universe throughout this work.
\subsection{Braneworld models}
Derived from the Braneworld theory, the Dvali-Gabadadze-Porrati model \citep{Dvali2000} alters gravity to replicate the cosmic acceleration, eliminating the need for Dark Energy (DE). In this model, we exist on a 4-D membrane within a higher-dimensional spacetime. Furthermore, gravity leaks into the bulk at large scales, leading to the accelerated expansion of the Universe \citep{Li2013}. In this subsection, we focus on two models: DGP and $\alpha$DE.

\subsubsection{Dvali-Gabadadze-Porrati model}
For DGP model, the Friedmann equation is governed by
\begin{equation}
E(z)^{2}-\frac{E(z)}{r_{c}}=\Omega_{m}(1+z)^{3},
\end{equation}
where $r_{c} = 1/[H_{0}(1-\Omega_{m})]$ is the crossover scale and $\Omega_{m}$ is the matter density at present. We can
directly rewrite the above equation and get the expansion
rate
\begin{equation} \label{Eq.2}
H(z)^2=H_0^2\left(\sqrt{\Omega_{m}(1+z)^{3}+\Omega_{rc}}+\sqrt{\Omega_{rc}} \right)^2,
\end{equation}
where  $\Omega_{rc}= 1/(4r_{c}^{2} H_{0}^{2})$ is
related to the cosmological scale. Setting $z = 0$ in Eq. (\ref{Eq.2}),
the normalization condition can be obtained
\begin{equation}
\Omega_{rc}=\frac{(1-\Omega_{m})^{2}}{4},
\end{equation}
and there are two model parameters $\hat{p} = (\Omega_{m}, H_{0})$ to be constrained.

\subsubsection{$\alpha$ Dark Energy model}
The $\alpha$DE is a phenomenological extension of DGP \citep{Dvali2003}. It was proposed by Dvali and Turner. By introducing a parameter q, it enables interpolation between the DGP and the $\Lambda$CDM. In this model, the Friedmann equation is modified to
\begin{equation}
E(z)^{2}-\frac{E(z)^{q}}{r_{c}^{2-q}}=\Omega_{m}(1+z)^{3},
\end{equation}
where $q$ is a phenomenological parameter, and $r_{c} = (1 -\Omega_{m})^{1/(q-2)}H_{0}^{-1}$.
According to this Friedmann equation, E(z) is determined by the following equation:
\begin{equation}
H(z)^{2}=H_0^2\left(\Omega_{m}(1+z)^{3}+\frac{H(z)^{q}}{H_0^{2-q}}(1-\Omega_{m})\right),
\end{equation}
when $q = 1$, it collapses to the DGP, when $q = 0$, it
collapses to $\Lambda$CDM. And there are three free parameters
 $\hat{p} = (\Omega_{m},q,H_{0})$ to be constrained.

\subsection{Chaplygin Gas models}
The Chaplygin Gas model \citep{Kamenshchik2001} is a fluid model that is often viewed as arising from the d-brane theory. As one of the candidates for dark energy models, the Chaplygin Gas model has been proposed to explain the cosmic acceleration \citep{Bento2002,Biesiada2005,Malekjani2011}. In this model, dark energy and dark matter are unified through an exotic equation of state. The universe is filled with the so-called Chaplygin gas, which is a perfect fluid characterized by the equation of state $p=-A/\rho$. However, the original Chaplygin Gas model has been excluded by observational data \citep{Davis2007}.
Therefore, we consider two other models: the Generalized Chaplygin Gas (GCG) model \citep{Bento2002} and the New Generalized Chaplygin Gas (NGCG) model \citep{Zhang2006}.

\subsubsection{Generalized Chaplygin Gas model}
The exotic equation of state of GCG \citep{Bento2002} can be expressed as:
\begin{equation}
p_{GCG}=-\frac{A}{\rho_{GCG}^{\xi}},
\end{equation}
where $A$ is a positive constant. Then, we can get the energy
density of GCG:
\begin{equation}
\rho_{GCG}=\rho_{0GCG} \left[A_{s}+(1-A_{s})a^{-3(1+\xi)} \right]^{\frac{1}{1+\xi}},
\end{equation}
where  $A_{s} = A/\rho_{0GCG}^{1+\alpha}$ is
the present energy density of the GCG. Note that GCG
behaves like a dust matter if $A_{s} = 0$, and GCG behaves like
a cosmological constant if $A_{s} = 1$. Considering a universe
with GCG, baryon, and radiation, we have
\begin{equation}
\begin{split}
H(z)^{2}=&H_0^2\bigg(\Omega_{b}(1+z)^{3}+\\
&(1-\Omega_{b})\left[A_{s}+(1-A_{s})(1+z)^{3(1+\xi)}\right]^{\frac{1}{1+\xi}}\bigg),
\end{split}
\end{equation}
where $\Omega_{b}$ is the present density parameter of the baryonic
matter. We adopt $100\Omega_{b} h^{2} = 2.166 \pm 0.015 \pm 0.011$ with
$h = H_{0}/100$ as usual and in the uncertainty budget first term
is associated with the deuterium abundance measurement
and the second one$-$with the Big Bang Nucleosynthesis
(BBN) calculation used to get $\Omega_{b0}$ \citep{Cooke2018}. Since
the parameter $A_{s}$ can be expressed by the effective total
matter density $\Omega_{m}$ and the $\xi$ parameter
\begin{equation}
A_{s}=1- \left(\frac{\Omega_{m}-\Omega_{b}}{1-\Omega_{b}}\right)^{1+\xi},
\end{equation}
there are three free parameters $\hat{p} = (\Omega_{m}, \xi, H_{0})$ in this model.

\subsubsection{New Generalized Chaplygin Gas model}
The GCG model can actually be viewed as an interacting model between vacuum energy and cold dark matter. If one wishes to further extend this model, a natural idea is to replace the vacuum energy with a more dynamic dark energy. Thus, the NGCG model was proposed \citep{Zhang2006}. The equation of state for the NGCG fluid is given by
\begin{equation}
p_{NGCG}=-\frac{\widetilde{A}(a)}{\rho_{NGCG}^{\xi}},
\end{equation}
where $\widetilde{A}(a)$ is a function of the scale factor $a$, and $\xi$ is a free
parameter. The energy density of NGCG can be expressed as
\begin{equation}
\rho_{NGCG}=\left[Aa^{-3(1+\xi)(1+\zeta)}+Ba^{-3(1+\xi)}\right]^{\frac{1}{1+\xi}},
\end{equation}
where A and B are positive constants. The form of the function $\widetilde{A}(a)$ can be determined to be
\begin{equation}
\widetilde{A}(a)=-\zeta Aa^{-3(1+\xi)(1+\zeta)},
\end{equation}
NGCG reduces to GCG if $\zeta = -1$, reduces to $\omega$CDM if
$\xi = 0$, and reduces to $\Lambda$CDM if ($\xi = 0, \zeta = -1$). In a flat
universe, we have
\begin{equation}
\begin{split}
&H(z)^{2}=H_0^2 \big(\Omega_{b}(1+z)^{3} +\\
&(1-\Omega_{b})(1+z)^{3}[1-\frac{\Omega_{de}}{1-\Omega_{b}}
(1-(1+z)^{3\zeta(1+\xi)})]^{\frac{1}{1+\xi}}\big),
\end{split}
\end{equation}
here, we set $\Omega_{de} = 1-\Omega_{m}$ and obtain the value of $\Omega_{b}$ from $100\Omega_{b} h^{2} = 2.166 \pm 0.015 \pm 0.011$ with $H_{0} =100h$. There are four free parameters
$\hat{p} = (\Omega_{m}, \xi,\zeta, H_{0})$ in this model.

\subsection{Emergent Dark Energy Models}
In this study, we investigate two models of Emergent Dark Energy: the Phenomenologically Emergent Dark Energy (PEDE) model \citep{Li2019}, and the Generalised Emergent Dark Energy (GEDE) model \citep{Li2020}.
They used these model to fit the latest astronomical observation data and explored its impact on solving the ``Hubble constant'' problem in current cosmology. By comparing the fitting results under different models, They found that these models have significant potential to alleviate the tension observed in the Hubble constant.

\subsubsection{Phenomenologically Emergent Dark Energy model}
The PEDE model \citep{Li2019} is a simple (zero degree of freedom) but radical phenomenological model of dark energy with symmetrical behavior around the current time, where dark energy and matter densities are comparable. The Friedmann equation for the PEDE model can be expressed as:
\begin{equation}
H(z)^{2}=H_{0}^{2}\left[\Omega_{m}(1+z)^{3}+\widetilde{\Omega}_{\rm{DE}}\right],
\end{equation}
where $\Omega_{m}$ is the matter density at present time and $\widetilde{\Omega}_{\rm{DE}}(z)$
can be expressed as:
\begin{equation}
\widetilde{\Omega}_{\rm{DE}}(z)=\Omega_{\rm{DE},0}\times \exp\left[3\int_{0}^{z}\frac{1+\omega(z')}{1+z'}dz'
\right],
\end{equation}
where $\omega(z) = p_{\rm{DE}}/\rho_{\rm{DE}}$ is the equation of state of Dark
Energy.

Here, we introduce the PEDE model in which the dark energy density has the following form:
\begin{equation}
\widetilde{\Omega}_{\rm{DE}}(z)=\Omega_{\rm{DE},0}\times[1-\tanh(\log_{10}(1+z))],
\end{equation}
where $\Omega_{\rm{DE},0} = 1-\Omega_{0m}$ and $1+z = 1/a$ where $a$ is the scale
factor. This dark energy model has no degree of freedom and
we can derive its equation of state following:
\begin{equation}
\omega(z)=-\frac{1}{3\ln(10)}\times\left(1+\tanh[\log_{10}(1+z)]\right)-1,
\end{equation}
there are two free parameters $\hat{p} = (\Omega_{m}, H_{0})$ in this model.
\subsubsection{Generalised Emergent Dark Energy model}
Subsequently, Li et al. proposed a generalised parameterization form for the PEDE model \citep{Li2020} named as GEDE model. This generalised parametric form incorporates two parameters to describe the properties of dark energy evolution: a free parameter $\Delta$ that characterizes the slope of dark energy density evolution, and a parameter  $z_{t}$ that determines the transition redshift where dark energy density equals matter density. The GEDE model offers flexibility, as it can encompass both the $\Lambda$CDM model and the PEDE model as two special cases, with $\Delta = 0$ and $\Delta = 1$. respectively.
 The Hubble parameter in this model can be expressed as

\begin{equation}
H(z)^{2}=H_{0}^{2}\left[\Omega_{m}(1+z)^{3}+\widetilde{\Omega}_{\rm{DE}}+\Omega_{\rm{R},0}(1+z)^{4}\right],
\end{equation}
$\Omega_{m}$and $\Omega_{R,0}$ is the current matter density and radiation
density, respectively. Here  $\widetilde{\Omega}_{\rm{DE}}(a)$ is defined as
\begin{equation}
\widetilde{\Omega}_{\rm{DE}}(a)=\Omega_{\rm{DE}}(a)\times \frac{H_{a}^{2}}{H_{0}^{2}},
\end{equation}
In GEDE model, the evolution for dark energy density has the following form:
\begin{equation}
\widetilde{\Omega}_{\rm{DE}}(z)=\Omega_{\rm{DE},0}\frac{1-\tanh\left(\Delta\times \log_{10}(\frac{1+z}{1+z_{t}})\right)}{1+\tanh(\Delta\times\log_{10}(1+z_{t}))},
\end{equation}
where $\Omega_{\rm{DE},0}=1-\Omega_{m}$ and transition redshift $z_{t}$
can be derived by the condition of $\widetilde{\Omega}_{\rm{DE}}(z_{t})=\Omega_{m}(1+z_{t})^{3}$. From the above formulas, we can get the expression about $\Omega_{m}$.

\subsection{cosmological models with torsion}

The Einstein-Cartan (EC) theory with the presence of space-time torsion has gained a lot of attention, where torsion acts a new dark source of torsionless Riemannian gravity and drives the expansion of Universe.
This  theory based on a simple physical intuition, that torsion effect is regarded as a macroscopic manifestation of the intrinsic angular momentum (spin) of matter  (e.g. see \cite{BH} for a recent review).
Recently, Pereira et al. introduced a torsion model that solely considers baryonic matter and a cosmological term \citep{Pereira2022}. The interaction between torsion and baryonic as well as cosmological constant terms naturally leads to an effective contribution from dark matter. Additionally, the cosmological constant term is modified due to the torsion coupling, resulting in an effective cosmological constant.
When torsion is present, the analogues of the Friedmann and Raychaudhuri equations in a space-time with a non-vanishing cosmological constant (i.e. when $ \Lambda\neq0$) assume the following form \citep{Kranas2019}:
\begin{equation} \label{eq.21}
\left(\frac{\dot{a}}{a}\right)^{2}=\frac{1}{3}\kappa\rho+\frac{1}{3}\Lambda-4\phi^{2}
-4\left(\frac{\dot{a}}{a}\right)\phi,
\end{equation}
and
\begin{equation}
\frac{\ddot{a}}{a}=-\frac{1}{6}\kappa(\rho+3p)+\frac{1}{3}\Lambda-2\dot{\phi}-
2\left(\frac{\dot{a}}{a}\right)\phi,
\end{equation}
where $\rho$ and $p$ are the energy density and pressure of matter, with the definition of the equation of state $\omega = p/\rho$, the continuity equation will be the following form:
\begin{equation}
\dot{\rho}+3(1+\omega)H\rho+2(1+3\omega)\phi\rho=4\phi\Lambda\kappa^{-1}
\end{equation}
where $H = \dot{a}/a$ is defined as the Hubble parameter. According to whether there is dark energy, it is divided into two models, the torsion model with dark energy and the torsion model without dark energy.

\subsubsection{Case I: vanishing cosmological constant}
In the case of the disappearance of cosmological constants, if assuming a empty space $(p = \Lambda= 0)$, which means that our universe only contains the matter component, and the accelerating expansion of the universe is driven by torsion. The Eq. (\ref{eq.21}) will become $\varphi(t) =-H(t)/2$, which means the torsion function $\varphi(t)$ depends on Hubble parameter $H$.  Thus, we adopt a simple parameterized form $\varphi(t) =-\alpha H(t)/2$, then the Friedmann equation becomes
\begin{equation}
H^{2}=\frac{\kappa}{3}\rho_{m}+4\alpha H^{2}-4\alpha^{2}H^{2}.
\end{equation}
This equation clearly shows that torsion contributes to the total effective energy-density of
the system. Similar to the FLRW Universe model, the density parameter $\Omega$ can also be
introduced for a homogeneous and isotropic model with torsion $\Omega_{m}+\Omega_{\alpha}=1$,
where the matter density parameter $\Omega_{m}=\frac{\kappa\rho_{m}}{3H^{2}}$ and the torsion density parameter $\Omega_{\alpha}=4\alpha-4\alpha^{2}$, then, we can get
\begin{equation}
H(z)^{2}=\frac{H_0^2}{(1-2\alpha)^{2}}(\Omega_{m}(1+z)^{(3-2\alpha)}),
\end{equation}
there are two free parameters $\hat{p} = (\Omega_{m},  H_{0})$ in this model.

\subsubsection{ Case II: non-vanishing cosmological constant}

For $\varphi(t) = -\alpha H(t)$ with $\Lambda\neq 0$ and $\omega = 0$, which means that  the accelerating expansion of the universe is driven by the cosmological constant plus torsion, one could obtain the Friedmann equation as
\begin{equation}
H^{2}=\kappa\rho_{m}+\frac{\Lambda}{3}+4\alpha H^{2}-4\alpha^{2}H^{2}.
\end{equation}
Similar to Case I, we introduce the density parameter for cosmological constant $\Omega_{\Lambda}=\frac{\Lambda}{3 H^{2}}$. For convenience, we also denote $ \Omega_{\Lambda}$ as the present value of cosmological constant density parameter. Based on the conservation equation of $\Omega_{m}+\Omega_{\Lambda}+\Omega_{\alpha}=1$, the Friedmann equation can be written as
\begin{equation}
\begin{split}
&H(z)^{2}=\frac{H_0^2}{(1-2\alpha)^{2}}\\
&\left[\left(\Omega_{m}+\frac{4\alpha}{(3-2\alpha)}\Omega_{\Lambda}
\right)(1+z)^{3-2\alpha}+\Omega_{\Lambda}\left(1-\frac{4\alpha}{3-2\alpha}\right)\right],
\end{split}
\end{equation}
there are three free parameters $\hat{p} = (\Omega_{m}, \alpha, H_{0})$ in this model.

\section{Observational data and METHODS}

In this section, we present the details of deriving observational
constraints on the cosmological models from X-ray and UV measurements of quasars and BAO
measurements.

\subsection{High-redshift observations of quasars and BAO}

Quasars have considerable potential to be used as useful cosmological probes. Although the extreme variability in their luminosity and high observed dispersion, many efforts have been made to standardize quasars as probes based on their own properties.  Recently, \citep{Risaliti2019} attempted to use quasars as standard candles
by using the nonlinear relation between their intrinsic UV and the X-ray luminosities. So far, the largest quasar sample with both X-ray and UV observations consists of $\sim12, 000$ objects, assembled by combining several different samples in \citep{Lusso2020}. Most of the quasars in the parent sample are from XMM-Newton
\citep{Nardini2019}, XMM-XLL sample \citep{Menzel2016} and SDSS \citep{Mingo2016}. Then several filtering steps were applied to reduce the systematic effects and 2421 quasars in the redshift range $0.009<z<7.5$ were left in the final cleaned sample \citep{Lusso2020}. The scatter plot of the final quasar sample is shown in Fig. 1.

The relation between the X-ray and UV luminosities is usually
parameterized as \citep{Risaliti2019}
\begin{equation}
\log(L_{X})=\gamma \log(L_{UV})+\beta,
\end{equation}
where $L_{X}$ and $L_{UV}$ are the rest-frame monochromatic luminosities
at $2$ keV and 2500 {\AA}, and $\log = \log_{10}$, the slope $-\gamma$ along with the intercept $-\beta$ are two free parameters.
Applying the flux-luminosity relation of $F=L/4\pi D_{L}^{2}$, the UV and X-ray luminosities can be replaced by the observed fluxes:
\begin{equation} \label{eq.28}
\log(F_{X})=\gamma\log(F_{UV})+(2\gamma-2)\log(D_{L}H_0)+\beta',
\end{equation}
where $\beta'=\beta-(2\gamma-2)\log(H_0)+(\gamma-1)\log4\pi$,  $F_{X}$ and $F_{UV}$ are the X-ray and UV fluxes, respectively. Here $D_{L}$ is the luminosity distance, which indicates
such kind of quasar measurements can be used to calibrate
them as standard candles. Theoretically, $D_{L}$ is determined
by the redshift $z$ and cosmological parameters $\hat{p}$ in a specific
model:
\begin{equation}
D_{L}(z,{\hat p})=\frac{c(1+z)}{H_{0}}\int_{0}^{z}\frac{d
z'}{E(z')}.
\end{equation}
In order to constrain cosmological
parameters ${\hat p}$ through the measurements of quasar X-ray and UV
fluxes, we compare the observed X-ray fluxes with the predicted
X-ray fluxes calculated with Eq. (\ref{eq.28}) at the same redshift. Then,
the best-fitted parameter values and respective uncertainties for
each cosmological model are determined by minimizing the $\chi^{2}$ objective function, defined by the log-likelihood \citep{Risaliti2019}
\begin{equation}
\chi^{2}_{QSO}=\sum_{i=1}^{2421}\left [\frac{[\log(F_{X,i}^{obs})-\log(F_{X,i}^{th})]^{2}}{s_{i}^{2}}+\ln(2\pi s_{i}^{2}) \right ],
\end{equation}
where $s_{i}^{2}=\sigma_{i}^{2}+\delta^{2}$, and $\sigma_{i}$ is the measurement error on $F_{X,i}^{obs}$. In addition to the cosmological model parameters, three more free parameters are fitted: $\gamma$, $\beta$ representing the X-UV relation and $\delta$ representing the global intrinsic dispersion. More recently, \citep{Li2021} presented a model-independent approach to calibrate this sample with latest SN Ia observations, and estimated that $\beta=7.735\pm0.244$, $\gamma=0.649\pm0.007$, and $\delta=0.235\pm0.04$. We will use their calibrated parameters in this work.

\begin{figure}
    \includegraphics[width=\columnwidth]{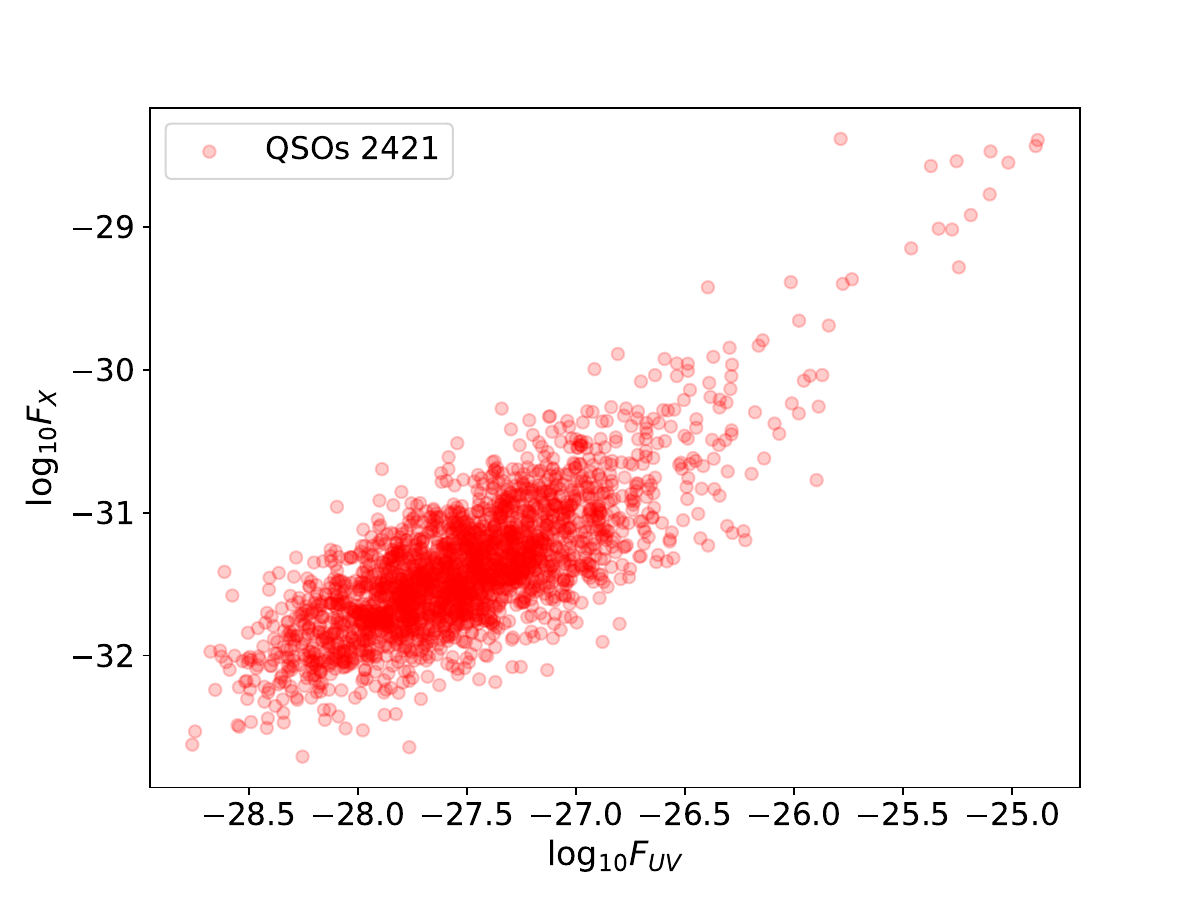}
    \caption{The scatter plot of UV and X-ray fluxes of 2421 quasars}
\end{figure}
However, considering the presence of high intrinsic dispersion of this quasar sample and the degeneracy between cosmological model parameters. The inclusion of both the transverse BAO measurements obtained without assuming the fiducial cosmological model (commonly known as 2D-BAO) and the line-of-sight and transverse BAO measurements from the spectroscopic survey (commonly known as 3D-BAO) aids in constraining the parameters of these cosmological models.

The transverse BAO measurements consist of 15 $\theta_{BAO}(z)$ values obtained from the public data releases (DR) of the Sloan Digital Sky Survey(SDSS) which are summarized in Table 1 of \citet{Nunes2020}. The BAO angular scale $\theta_{BAO}(z)$ is related to the angular diameter distance as follows
\begin{equation}
	\theta^{th}_{BAO}(z)=\frac{r_d}{(1+z)D_A(z)}
\end{equation}
where $r_d$ is the comoving sound horizon at the baryon drag epoch. The calibration of $r_d$ enables the utilization of the BAO angular scale $\theta_{BAO}(z)$ to establish a precise correlation between angular diameter distance and redshift. For the sound horizon at the drag epoch,  we adopt the value of $r_d=147.59$ Mpc \cite{Nunes2020} reported from Planck plus transversal BAO data as a prior value in our work. Then the best-fitted parameter values are determined by corresponding log-likelihood function
\begin{equation}
	\chi_{2D-BAO}^{2}({\hat p})=\sum_{i=1}^{15}\frac{\left [\theta^{th}
		_{BAO}(z_{i};{\hat p})-\theta^{obs}_{BAO}(z_{i})\right ]^{2}}{\sigma^{2}_
		{BAO(z_{i})}}.
\end{equation}

For the use of 3D-BAO data, we follow the approach carried out in \citep{Alam2021}. The measurements obtained from the completed Sloan Digital Sky Survey (SDSS) Main Galaxy Sample (MGS), the Baryon Oscillation Spectroscopic Survey(BOSS) Data Release 12 (DR12) galaxies, the extended Baryon Oscillation Spectroscopic Survey (eBOSS) galaxies and quasars, the auto-correlation of the Ly$\alpha$ forest and its cross-correlation with quasars from BOSS and eBOSS, are summarized as BAO-only measurements in Table 3 of \citet{Alam2021}. The relevant quantities include the Hubble distance $D_H(z)$, the transverse comoving distance $D_M(z)$ and the spherically averaged angular-diameter distance $D_V(z)$.

The Hubble distance along the line-of-sight direction is determined by the speed of light and the Hubble parameter
\begin{equation}
	D_H(z)=\frac{c}{H(z)}.
\end{equation}
The comoving distance in the transverse direction can be simplified in the case of a spatially flat universe as
\begin{equation}
	D_M(z)=c\int_{0}^{z}\frac{dz'}{H(z')}.
\end{equation}
The spherically averaged angular-diameter distance can be readily utilized by taking into account both the line-of-sight and transverse directions as
\begin{equation}
	D_V(z)=[zD^2_M(z)D_H(z)]^{1/3}.
\end{equation}

Considering the dependence of the comoving sound horizon $r_d$ at baryon drag epoch , expressed as  $D_H(z)/r_d$, $D_M(z)/r_d$, and $D_V(z)/r_d$ more directly. The expressions for these three BAO distances are represented by the quantities $A_{obs}(z_i)$, $\sigma_{A_i}$, and $A_{th}(z_i)$, which correspond to the measurements, uncertainties, and theoretical expression, respectively.For the uncorrelated BAO measurements, the corresponding log-likelihood function is given by
\begin{equation}
\chi_{3D-BAO}^{2}({\hat p})=\sum_{i=1}^{3}\frac{\left [ A
_{th}(z_{i};{\hat p})-A _{obs}(z_{i})\right ]^{2}}{\sigma
(z_{i})^{2}}.
\end{equation}
For the correlated BAO measurements, the log-likelihood function of correlated BAO measurements takes the for
\begin{equation}
\chi_{3D-BAO}^{2}({\hat p})=\left [ A _{th}({\hat p})-A _{obs}\right
]^{T}C^{-1}\left [ A _{th}({\hat p})-A _{obs}\right ],
\end{equation}
where $C^{-1}$ is the inverse covariance matrix.

\subsection{MCMC analysis}

In order to determine the cosmological parameters $p$ in different models, we use MCMC method to achieve our purpose, implemented in the \emph{emcee} package
\footnote{https://pypi.python.org/pypi/emcee} in Python
\citep{Foreman-Mackey2013}. Using the $\chi^{2}$ objective function defined above, the total  $\chi^{2}_{tot}$ function of the above combined analysis is expressed as
\begin{equation}
\chi^{2}_{tot}=\chi^{2}_{QSO}+\chi^{2}_{BAO},
\end{equation}
we can also write the likelihood function as
\begin{equation}
\mathcal{L}({\hat p})=\exp^{-\frac{\chi^{2}_{tot}}{2}}.
\end{equation}
Then it is essential to determine which model is most preferred by the observational measurements and carry out a good comparison between these different models. Out of possible model selection techniques, we will use the Akaike Information Criterion (AIC) \citep{Akaike1974}
\begin{equation}
\rm AIC=\chi_{min}^{2}+2k,
\end{equation}
as well as the Bayesian Information Criterion (BIC) \citep{Schwarz1978}
\begin{equation}
\rm BIC=\chi_{min}^{2}+k\ln N,
\end{equation}
where $\chi_{min}^{2}=-2\ln\mathcal{L}_{max}$, $k$ is the number of free parameters
in the model and $N$ represents the number of data points.

\section{Results and discussions}

\begin{figure}
\begin{center}
    \includegraphics[width=0.8\columnwidth]{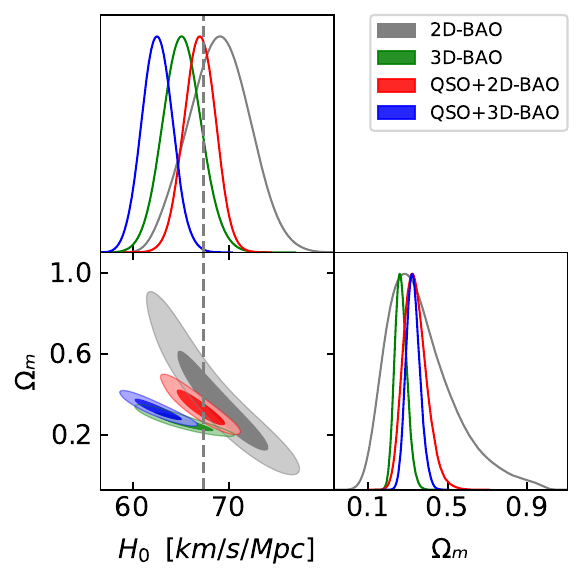}
    \caption{The 1D probability distributions and 2D contours with
        $1\sigma$ and $2\sigma$ confidence levels for the DGP model obtained
         from the 2-D and 3-D BAO and combinations of QSO+BAO datasets, respectively. The black dashed line
        represents the $H_0=67.4$ $km/s/Mpc$ reported by Planck 2018 results.}
\end{center}
\end{figure}

\begin{figure}
\begin{center}
    \includegraphics[width=0.9\columnwidth]{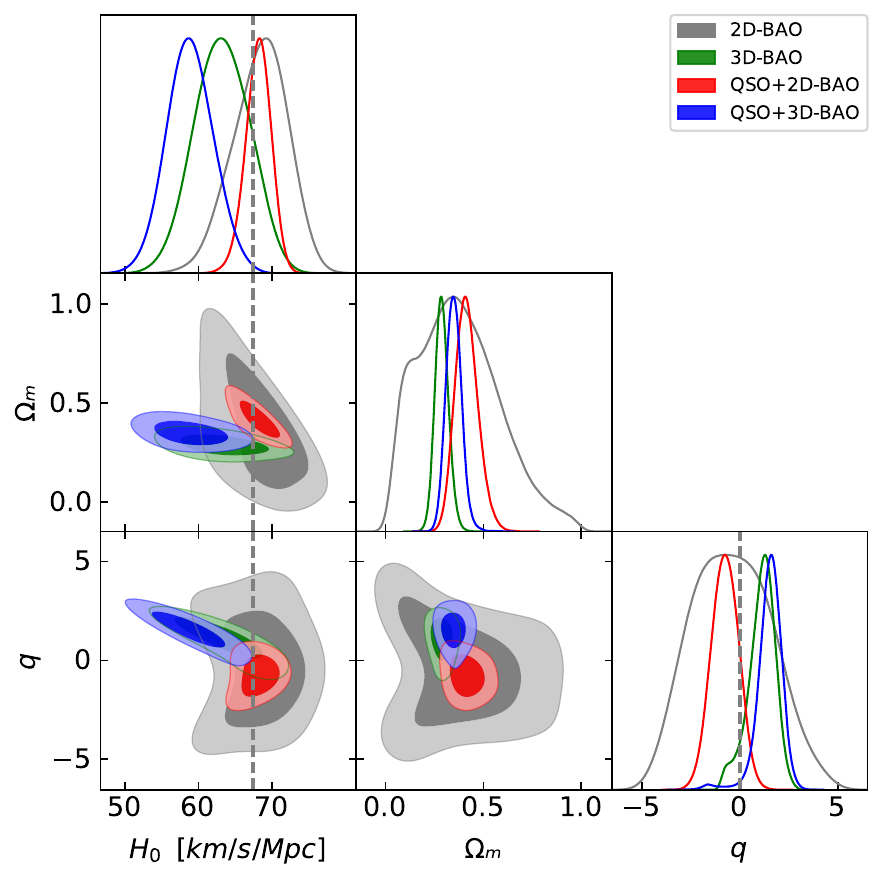}
    \caption{Similar to Figure 1, but for the $\alpha$DE model. The black dashed line
        represents the $\Lambda$CDM model corresponding to q = 0.}
\end{center}
\end{figure}

In this section, we give the results of the cosmological models listed in Section 2, obtained using two special BAO compilations (2D-BAO and 3D-BAO) and the combinations of BAO and QSO datasets (QSO+2D-BAO and QSO+3D-BAO), respectively. The distinction between the outcomes obtained from the inclusion of QSO data and BAO data alone is denoted as BAO and QSO+BAO, respectively, without explicitly distinguishing the BAO datasetin part text of this discussion section. In order to get a better comparison, the corresponding results for the concordance $\Lambda$CDM model is also displayed. The 1D probability distributions and 2D contours with $1\sigma$ and $2\sigma$ confidence levels, as well as the best-fit value with 1$\sigma$ uncertainty for each model are shown in Figs. 2-9 and reported in Table I.

\subsection{Observational Constraints on Braneworld models}

As can be seen from Table I, the best-fit values of $\Omega_{m}$ in the DGP is $\Omega_{m}=0.32_{-0.03}^{+0.04}$ within $68.3\%$ confidence level by using the most comprehensive of QSO + 3D-BAO datasets currently accessible, which agree well with the recent Planck 2018 results: $\Omega _{m}=0.315\pm 0.007$ \citep{Planck2018} and SNe Ia+BAO+CMB+observational Hubble parameter
(OHD): $\Omega _{m}=0.305\pm 0.015$ \citep{Shi2012}. For the 3D-BAO data, the best-fit matter density parameters for this DGP model is $\Omega_{m}=0.26_{-0.03}^{+0.03}$, which is significantly lower than the QSO + 3D-BAO mentioned above. The combination of 2D-BAO data yielded similar findings, however, a significant disparity was observed when compared to the results obtained from 3D-BAO data. It is worthwhile to mention that the matter density parameter $\Omega_{m}$ obtained by QSO tends to be higher than that from other cosmological probes, as was remarked in the previous works
of \citep{Risaliti2019,Khadka2020}. This suggests that the composition of the universe characterized by cosmological parameters can be comprehended differently through high-redshift quasars. The use of BAO data can effectively help QSOs constrain cosmological models, considering the constrained results especially on $\Omega_{m}$ and $H_{0}$. The best-fit values of $\Omega_{m}$ and $H_{0}$ for $\alpha$DE are noticeably higher than those of DGP, indicating a potential reason could be the presence of an additional parameter $q$ in $\alpha$DE. For the $\alpha$DE model, the
limit of $q = 1$ corresponds to the DGP model, and the limit of $q = 0$ corresponds to the $\Lambda$CDM model. It is evident that the optimal values of $q$ obtained from analyzing both 3D-BAO and 2D-BAO datasets in the context of $\alpha$DE favor the DGP and $\Lambda$CDM models respectively. We can also see that the BAO data alone can only provide a rather weak constraint on the parameter $q$ in $\alpha$DE, but the combined QSO+BAO data can constrain $q$ tightly. For the constrained results on $H_0$, the best-fit values of $H_0$ in the DGP and $\alpha$DE are both closed with recent Planck 2018 results when incorporating 3D-BAO measurements. In Fig. 2-3, for DGP and $\alpha$DE models, we can clearly see that the parameter degeneracy directions of BAO and BAO+QSO are almost identical, and by adding quasar data, the constrained parameters precision have been significantly improved.

\subsection{Observational Constraints on Chaplygin gas models}
The Table I presents the results of the best-fitted parameters for the GCG model. One can find large difference between the constraints on $\Omega_m$ and  $H_{0}$ obtained from the BAO and QSO+BAO datasets and the results derived from the 2D-BAO and 3D-BAO datasets exhibit significant disparities. In the context of GCG, it is worth noting that the parameter $\xi$ quantifies the deviation from the $\Lambda$CDM model and the original Chaplygin gas model. Consequently, $\Lambda$CDM is not consistent with GCG at the $1\sigma$ confidence level.

For the NGCG model, the estimated values of the cosmic parameters are presented in Table 1. It is observed that for the BAO and QSO+BAO datasets, the best-fit values of the parameters are in agreement with each other within the $2\sigma$ confidence level. It is worth noting that the matter density parameter $\Omega_m$ derived from such dataset  combinations except QSO+2D-BAO tends to be higher than that obtained from other cosmological probes, as previously reported in \cite{Liu2019c}. And the constraint results of $\xi \simeq 0$ and $\zeta\simeq-1$ from Table I indicate that the $\Lambda$CDM limit of these models is strongly favored by the current observations.
In terms of the constrained results on $H_0$, the best-fit values of $H_0$ in the NGCG models with 2D-BAO measurements align with recent Planck 2018 results. Additionally, due to the degeneracy between model parameters, the uncertainty of the parameters is quite large. This conclusion can be visually inferred from Figs. 4-5, emphasizing the significance of our work by highlighting the distinct roles played by BAO and quasar data in constraining cosmological model parameters. The inclusion of quasar data effectively breaks the degeneracy between model parameters.

\begin{figure}
    \includegraphics[width=0.9\columnwidth]{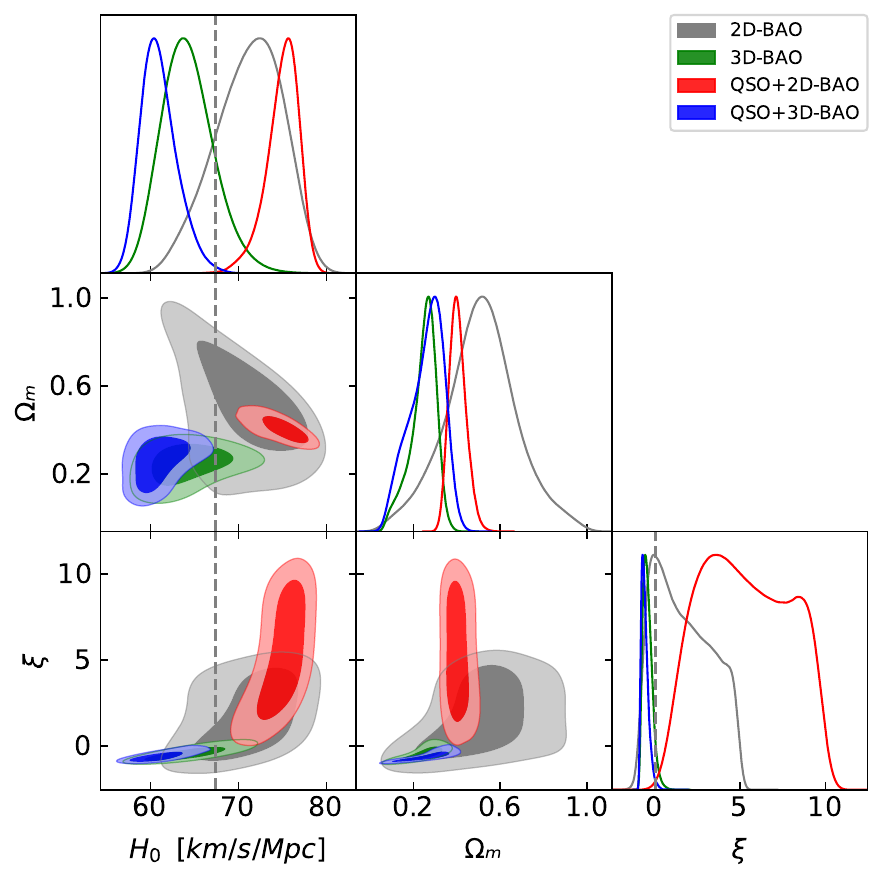}
    \caption{Similar to Figure 1, but for the GCG model. The black dashed line
        represents the $\Lambda$CDM model corresponding to $\xi$ = 0.}
\end{figure}

\begin{figure}
    \includegraphics[width=\columnwidth]{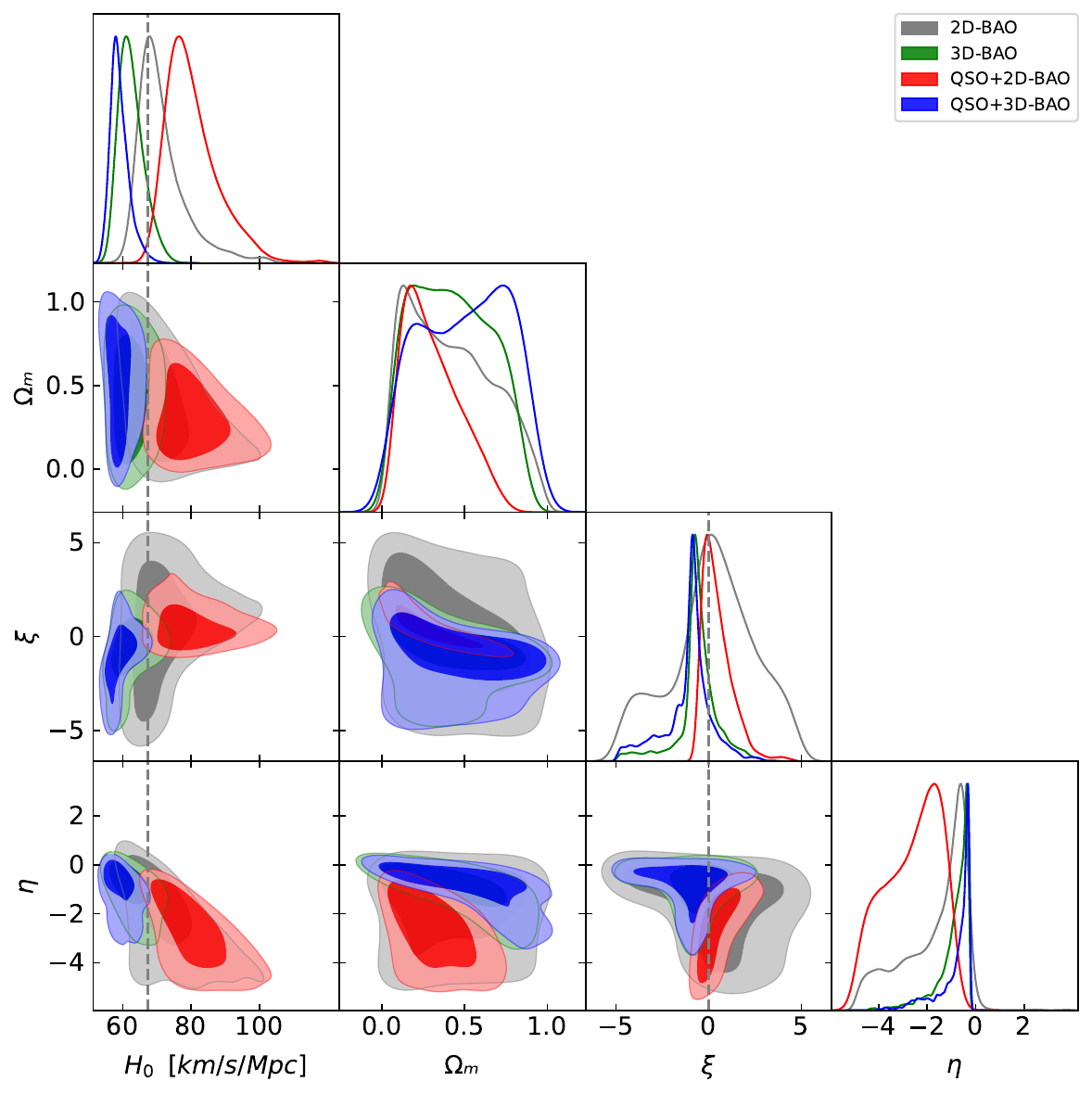}
    \caption{Similar to Figure 1, but for  the NGCG model. The black dashed line
        represents the $\Lambda$CDM model corresponding to $\xi$ = 0.}
\end{figure}

\subsection{Observational Constraints on Emergent Dark Energy Models}
The numerical results and 1$\sigma$ uncertainties of the constraints obtained using BAO alone and QSO+BAO datasets for the PEDE and GEDE models are summarized in Table I. The 1D probability distributions and 2D contours with $1\sigma$ and $2\sigma$ confidence levels are also illustrated in Fig. 6-7. For the $\Lambda$CDM model, $\Delta$ is fixed at zero, while for the PEDE model it is set to one. However, in the GEDE model, $\Delta$ is a free parameter. For the PEDE model, the best-fit value of $\Omega_{m}$, obtained using 3D-BAO datasets, is consistent with the results derived from Pantheon+BAO datasets \citep{Li2019} and the value of $H_0$ obtained using QSO+3D-BAO is consistent with the result obtained from Planck 2018 results. For the GEDE model, the best-fit values of $\Omega_{m}$ are $0.31^{+0.33}_{-0.38}$ by using 3D-BAO alone are consistent to the results obtained with the work \citep{Li2020}.
It is worth mentioning that the constrained results on $\Omega_{m}$ were calculated using the model parameters $\Delta$  and $z_{t}$, whose relationship is provided in the work \cite{Li2020}. Additionally, based on the GEDE model, our constraints on $H_0$ towards to a smaller value align more closely with recent Planck 2018 results. These results are $H_0=62.51_{-3.23}^{+3.65}$ $km/s/Mpc$ and $H_0=58.31_{-2.42}^{+3.19}$ $km/s/Mpc$  for 3D-BAO alone and QSO+3D-BAO datasets, respectively.
The best-fitting values of $\Delta$ from 2D-BAO and QSO+2D-BAO are consistent with zero within $1\sigma$ uncertainties, indicating that our results include the $\Lambda$CDM model but not the PEDE model. This is because, compared to the PEDE model, the GEDE model incorporates additional parameters, introducing greater uncertainty in the constrained results. The constrained results suggest that data may prefer simpler and more deterministic models to describe the behavior of the universe. In terms of PEDE and GEDE models, none of the model parameters break the degeneracy. However, the constrained parameter precision using QSO+BAO data is significantly higher than that using BAO data alone.

\begin{figure}
\begin{center}
    \includegraphics[width=0.8\columnwidth]{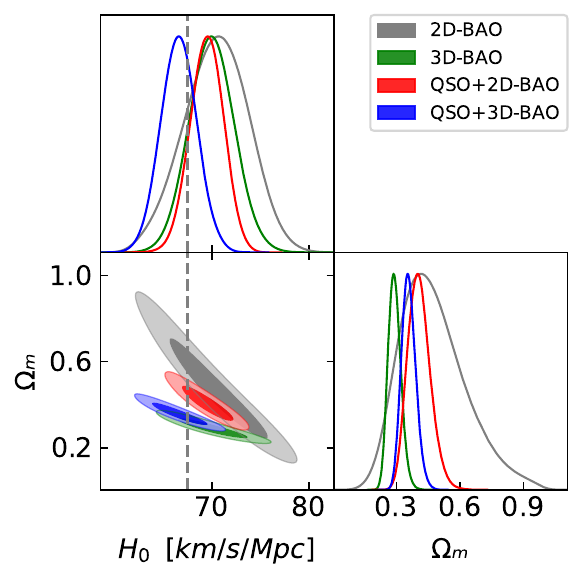}
    \caption{Similar to Figure 1, but for  the PEDE model.}
\end{center}
\end{figure}

\begin{figure}
\begin{center}
    \includegraphics[width=0.9\columnwidth]{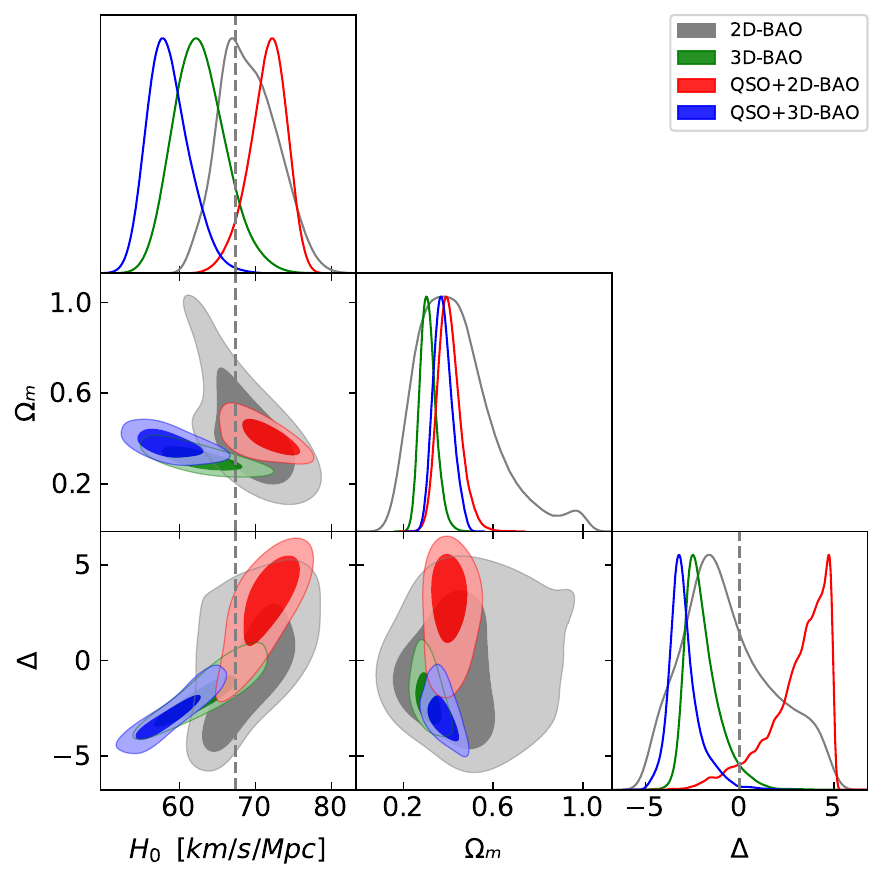}
    \caption{Similar to Figure 1, but for  the GEDE model. The black dashed line
        represents the $\Lambda$CDM model corresponding to $\Delta$ = 0.}
\end{center}
\end{figure}

\subsection{Observational Constraints on cosmological model with torsion}
\begin{figure}\renewcommand\arraystretch{0.5}
    \includegraphics[width=0.8\columnwidth]{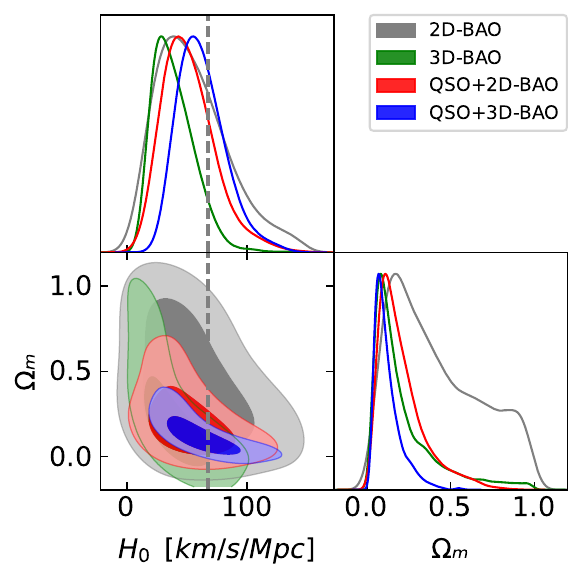}
    \caption{Similar to Figure 1, but for  the  CASE I model. }
\end{figure}

\begin{figure}\renewcommand\arraystretch{0.5}
    \includegraphics[width=0.9\columnwidth]{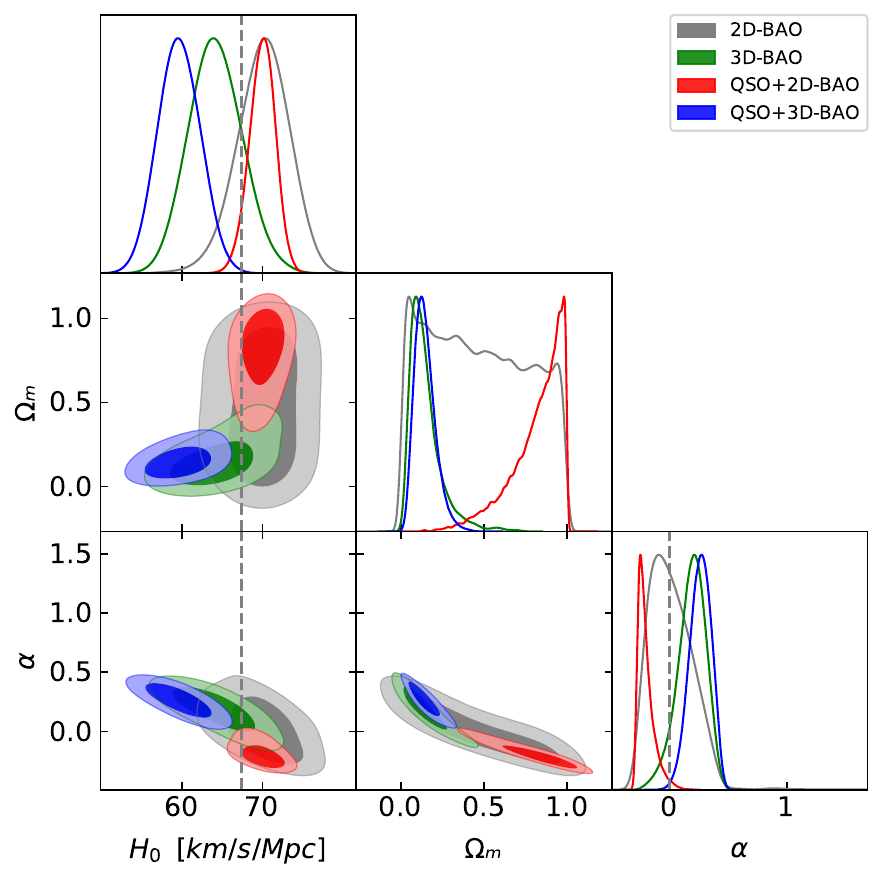}
    \caption{Similar to Figure 1, but for  the CASE II model. The black dashed line
        represents the $\Lambda$CDM model corresponding to $\alpha$= 0.}
\end{figure}

In Case I model, the torsion term leads to the accelerated expansion of the Universe without the need for a cosmological constant $\Lambda$. The Figure 8 shows the 1$\sigma$ and 2$\sigma$ contours for the parameters $H_0$ and $\Omega_m$. The best-fitting values of the Hubble constant from 3D-BAO data are $H_{0}=36.08_{-14.68}^{+20.36}$ $km/s/Mpc$ and $\Omega_{m}=0.15_{-0.09}^{+0.23}$. It is clear that 3D-BAO data alone are not sufficient to provide strong constraints on model parameters. However, when QSO+3D-BAO data are used, the values change to  $H_{0}=59.17_{-17.73}^{+22.36}$ $km/s/Mpc$ and $\Omega_{m}=0.10_{-0.05}^{+0.09}$.
The central value of $H_0$ also disagrees with the result inferred from Planck 2018 observation and local measurement by SH0ES collaboration. This conflict is likely due to a strong degeneracy between the Hubble constant and matter density. It is worth noting that the torsion parameter $\alpha$  does not introduce any extra degrees of freedom and is related to the density of matter parameter through the conservation condition $\Omega_{m}+4\alpha-4\alpha^2=1$. Therefore, we calculate $\alpha=0.37_{-0.03}^{+0.03}$ and $\alpha=0.34_{-0.03}^{+0.03}$ using 3D-BAO alone and QSO+3D-BAO datasets, respectively, which are consistent with the work \cite{2023JCAP...07..059L}. However, these results  conflict with previous results indicating a small value of $\alpha$ supported by low-redshift Hubble parameter and SN Ia observations \citep{Pereira2019,Pereira2022}.

In the context of Case II model, the torsion term and cosmological constant $\Lambda$ work in tandem to drive the accelerated expansion of the Universe. The Figure 9 shows the constraint results for $H_{0}$, $\Omega_{m}$, and $\alpha$ using high-redshift quasar and BAO observations. Table I summarizes the best-fit parameters, including their 1$\sigma$ uncertainties.
We obtain that $H_{0}=70.26_{-3.33}^{+3.09}$ $km/s/Mpc$ and $H_0=70.02^{+1.53}_{-1.67}$ $km/s/Mpc$  from 2D-BAO and QSO+3D-BAO data towards to a smaller value align with the Planck 2018 CMB observations when assuming a flat $\Lambda$CDM model.  Notably, the central values of $\alpha=-0.01^{+0.21}_{-0.17}$ from 2D-BAO alone align with the torsion's impact on primordial helium-4 abundance, where a narrow parameter range of  $-0.0058 <\alpha< +0.0194$ was reported in \cite{Kranas2019}.

\begin{table*} \label{Table.1}
\renewcommand\arraystretch{1}
\caption{Summary of the best-fit values with their 1$\sigma$
            uncertainties concerning the parameters of all considered
            models. The results are obtained from the 2-D and 3-D BAO and combinations of QSO+BAO datasets, respectively.}
\begin{center}
\setlength{\tabcolsep}{1mm}{
\begin{tabular}{c c c c c c}
\hline
\hline

Model & Data & $\Omega _{m}$  & $H_{0}$ [km/s/Mpc] \\
\hline

            $\Lambda$CDM
            & 2D-BAO & $0.43^{+0.19}_{-0.15}$& $70.05^{+3.08}_{-3.33}$\\
            & 3D-BAO & $0.29^{+0.03}_{-0.03}$& $68.03^{+2.13}_{-2.10}$\\
            & QSO+2D-BAO & $0.40^{+0.06}_{-0.05}$& $68.40^{+1.59}_{-1.61}$ \\
            & QSO+3D-BAO & $0.36^{+0.04}_{-0.03}$& $65.10^{+1.74}_{-1.69}$\\
            \hline

            Model & Data & $\Omega _{m}$  & $H_{0}$  [km/s/Mpc] \\
            \hline

             DGP
            & 2D-BAO & $0.34^{+0.20}_{-0.13}$ & $69.02^{+3.17}_{-3.17}$ \\
            & 3D-BAO & $0.26^{+0.03}_{-0.03}$ & $65.14^{+2.04}_{-1.96}$ \\
            & QSO+2D-BAO & $0.33^{+0.06}_{-0.05}$ & $66.98^{+1.61}_{-1.64}$\\
            & QSO+3D-BAO & $0.32^{+0.04}_{-0.03}$ & $62.64^{+1.57}_{-1.58}$ \\
            \hline

           Model & Data & $\Omega _{m}$  & $H_{0}$ [km/s/Mpc]  \\
            \hline

            PEDE
            & 2D-BAO & $0.46^{+0.17}_{-0.13}$ & $70.45^{+3.25}_{-3.46}$ \\
            & 3D-BAO & $0.29^{+0.03}_{-0.03}$ & $69.92^{+2.32}_{-2.29}$ \\
            & QSO+2D-BAO & $0.40^{+0.06}_{-0.05}$ & $69.46^{+1.69}_{-1.74}$\\
            & QSO+3D-BAO & $0.36^{+0.04}_{-0.03}$ & $66.47^{+1.91}_{-1.86}$\\
            \hline

           Model & Data & $\Omega _{m}$ & $q$  & $H_{0}$  [km/s/Mpc] \\
            \hline

           $\alpha$DE
            & 2D-BAO & $0.36^{+0.23}_{-0.21}$ & $-0.45^{+2.17}_{-2.10}$ & $68.61^{+3.47}_{-4.08}$ \\
            & 3D-BAO & $0.29^{+0.04}_{-0.03}$& $0.99^{+0.70}_{-2.58}$ & $63.24^{+3.83}_{-3.84}$ \\
            & QSO+2D-BAO & $0.41^{+0.06}_{-0.06}$& $-0.76^{+0.70}_{-0.70}$ & $68.18^{+1.66}_{-1.77}$ \\
            & QSO+3D-BAO & $0.35^{+0.04}_{-0.04}$& $1.56^{+0.52}_{-0.58}$ &$58.83^{+3.28}_{-3.09}$\\
            \hline
           Model & Data & $\Omega _{m}$ & $\xi$  & $H_{0}$  [km/s/Mpc] \\
            \hline

             GCG
            & 2D-BAO & $0.51^{+0.16}_{-0.16}$& $1.50^{+2.19}_{-1.65}$ &$71.47^{+3.62}_{-4.28}$ \\
            & 3D-BAO & $0.26^{+0.05}_{-0.07} $& $-0.49^{+0.31}_{-0.23}$ &$64.07^{+3.14}_{-2.75}$\\
            & QSO+2D-BAO & $0.40^{+0.04}_{-0.04}$ &$5.21^{+3.15}_{-2.71}$ &$75.25^{+1.52}_{-2.03}$\\
            & QSO+3D-BAO & $0.27^{+0.07}_{-0.10}$& $-0.64^{+0.24}_{-0.16}$ &$60.79^{+2.30}_{-1.81}$\\
            \hline

          Model & Data& $\Omega _{m}$ & $\alpha$  & $H_{0}$  [km/s/Mpc] \\
            \hline

            CASE I
            & 2D-BAO & $0.35^{+0.40}_{-0.22}$& $0.28^{+0.39}_{-0.15}$&$49.85_{-26.10}^{+34.62}$\\
            & 3D-BAO &$0.15^{+ 0.23}_{-0.09}$ &$0.37^{+ 0.03}_{-0.03}$ &$36.08_{-14.68}^{+20.36}$\\
            & QSO+2D-BAO &$0.17^{+0.17}_{-0.09}$ &$0.33^{+ 0.07}_{ -0.07}$ &$48.54_{-19.23}^{+ 23.65}$ \\
            & QSO+3D-BAO &$0.10^{+0.09}_{-0.05}$ &$0.34^{+ 0.03}_{-0.03}$ &$59.17_{-17.73}^{+ 22.36}$\\
            \hline

          Model & Data & $\Omega _{m}$ & $\alpha$  & $H_{0}$  [km/s/Mpc] \\
            \hline

           CASE II
            & 2D-BAO & $0.44^{+0.38}_{-0.31}$&$-0.01^{+0.21}_{-0.17}$&$70.26^{+3.09}_{-3.33}$ \\
            & 3D-BAO & $0.12^{+0.10}_{-0.06}$&$0.20^{+0.11}_{-0.13}$ &$64.10^{+3.44}_{-3.22}$\\
            & QSO+2D-BAO & $0.86^{+0.10}_{-0.20}$ & $-0.22^{+0.08}_{-0.04}$ &$70.02^{+1.53}_{-1.67}$\\
            & QSO+3D-BAO & $0.14^{+0.07}_{-0.05}$& $0.27^{+0.09}_{-0.10}$ &$59.63^{+2.70}_{-2.60}$\\
            \hline

           Model & Data& $\Omega _{m}$ & $\xi$ & $\zeta$ & $H_{0}$  [km/s/Mpc] \\
            \hline

           NGCG
            & 2D-BAO & $0.39^{+0.33}_{-0.26}$&$0.29^{+2.34}_{-2.61}$&$-1.20^{+0.73}_{-2.07}$&$69.61^{+8.04}_{-4.41}$ \\
            & 3D-BAO & $0.41^{+0.28}_{-0.26}$&$-0.62^{+0.85}_{-0.58}$&$-0.70^{+0.33}_{-0.75}$ &$61.90^{+4.25}_{-3.15}$\\
            & QSO+2D-BAO & $0.27^{+0.22}_{-0.15}$ &$0.28^{+0.97}_{-0.59}$&$-2.40^{+1.05}_{-1.61}$&$78.71^{+9.11}_{-5.60}$ \\
            & QSO+3D-BAO & $0.51^{+0.29}_{-0.35}$& $-0.90^{+0.78}_{-1.49}$&$-0.53^{+0.24}_{-0.98}$&$58.41^{+3.09}_{-2.00}$\\
            \hline

            Model & Data & $\Omega _{m}$ & $\Delta$ & $z_t$ & $H_{0}$  [km/s/Mpc] \\
            \hline

             GEDE
            & 2D-BAO & $0.41^{+0.19}_{-0.28} $& $0.15^{+0.44}_{-0.29}$&$-0.98^{+3.13}_{-1.98}$&$68.81^{+4.24}_{-3.42}$\\
            & 3D-BAO & $0.31^{+0.33}_{-0.38}$&$0.53^{+0.22}_{-0.15}$&$-2.16^{+1.09}_{-0.65}$&$ 62.51^{+3.65}_{-3.23}$ \\
            & QSO+2D-BAO & $0.40^{+0.45}_{-0.45}$ &$0.10^{+0.06}_{-0.05}$&$3.50^{+1.09}_{-2.01}$&$71.83^{+2.21}_{-2.70}$ \\
            & QSO+3D-BAO & $0.38^{+0.29}_{-0.36}$ &$0.47^{+0.31}_{-0.19}$&$-3.09^{+0.87}_{-0.55}$&$58.31^{+3.19}_{-2.42}$ \\
            \hline
            \hline

        \end{tabular}}

    \end{center}
\end{table*}

\begin{table*}\renewcommand\arraystretch{1.1}
\caption{Information theoretic model comparison. Minimum values of
            AIC, BIC, their differences are reported for the
            $\Lambda$CDM and each of  cosmological models considered. We give ranking of the favor degree of the model, the smaller the number represents the more favored by the observational data. }
    \begin{center}
        \setlength{\tabcolsep}{1.3mm}{
        \begin{tabular}{c c c c c c c c c c} 
            \hline
            \hline

            Dataset & Model & AIC & $\Delta$AIC &Favor Degree & Ranking  & BIC & $\Delta$BIC & Favor Degree &  Ranking  \\
            \hline

            QSO+2D-BAO & $\Lambda$CDM & 2406.42 & 5.44 & Excluded & 5 & 2418.02 & 1.73&Preferred & 2 \\
            & DGP & 2435.95 & 34.97 & Excluded& 8 & 2447.55 & 31.26& Excluded & 8  \\
            & $\alpha$DE & 2408.00 & 7.02 & Excluded& 6 & 2425.39 & 9.10& Excluded & 6  \\
            & PEDE& 2404.69  & 3.71& Slight   & 4 & 2416.29 & 0& Preferred & 1  \\
            & GEDE & 2403.65 & 2.67& Slight & 3 & 2421.04 & 4.75& Slight & 5  \\

            & GCG & 2400.98 & 0& Preferred& 1 & 2418.37 & 2.08 & Slight& 3 \\
            & NGCG & 2411.26 & 10.28& Excluded & 7 & 2434.45 & 18.16& Excluded & 7  \\

            & CASE I & 2457.76 & 56.78& Excluded & 9 & 2475.15 & 58.86 & Excluded& 9  \\
            & CASE II & 2402.39 & 1.41 & Preferred& 2 & 2419.78 & 3.49 & Slight& 4  \\

            \hline

           QSO+3D-BAO& $\Lambda$CDM & 2393.38& 6.76& Excluded& 8 & 2404.97& 4.78& Slight& 3  \\
            & DGP & 2388.60 & 1.98 & Preferred& 4 & 2400.19 & 0 & Preferred& 1  \\
            & $\alpha$DE & 2386.62 & 0 & Preferred& 1 & 2406.01 & 5.82 & Excluded& 4  \\
            & PEDE & 2396.45 & 9.83& Excluded &9 & 2408.04 & 7.85 & Excluded& 8  \\
            & GEDE & 2389.30 & 2.68 & Slight& 6 & 2406.69 & 6.50 & Excluded& 7  \\
            & GCG & 2386.65 & 0.03 & Preferred& 2 & 2406.04 & 5.85 & Excluded& 6  \\
            & NGCG & 2389.65 & 3.03& Slight &7 & 2412.83 & 12.64 & Excluded& 9  \\

            & CASE I & 2388.76 & 2.14 &Slight& 5 & 2406.15 & 5.96 & Excluded& 5  \\
            & CASE II & 2387.48 & 0.86 &Preferred& 3 & 2404.87 & 4.68&Slight & 2  \\

            \hline
            \hline

        \end{tabular}}

    \end{center}
\end{table*}
\subsection{MODEL COMPARISON}
\begin{figure}
    \includegraphics[width=9.2cm,height=7.3cm]{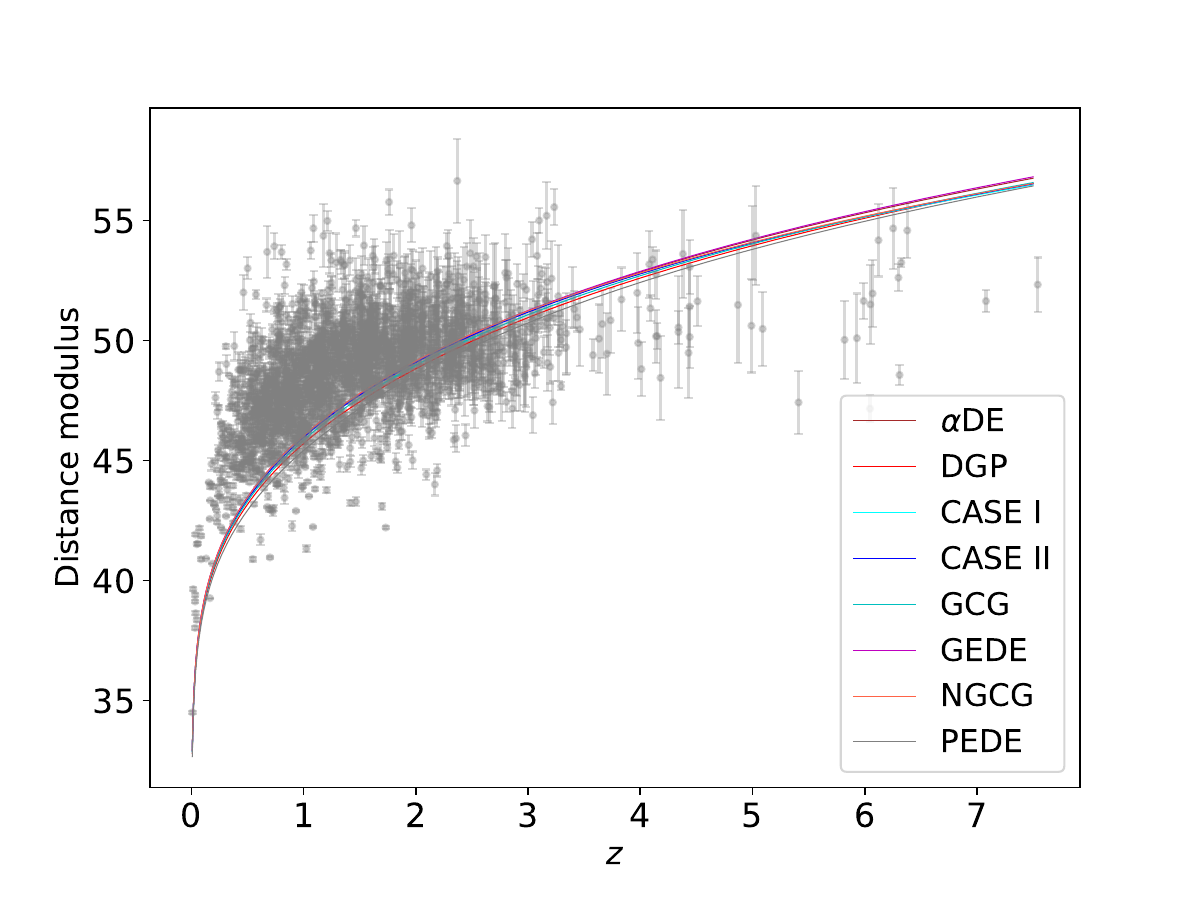}
    \caption{The Hubble diagram for the 2421 newest large sample of QSO X-ray and UV flux measurements. The solid lines represent the theoretical predictions of all models with the best-fit values obtained using quasar data.}
\end{figure}

We compare the models and discuss how strongly are they supported by the observational data sets in this section. In the Table II, one can find the summary of the information theoretical model selection criteria applied to different models from QSO+2D-BAO and QSO+3D-BAO data sets. There is an accepted interpretation of the magnitude of the difference $\Delta=\rm{AIC_{2}}-\rm{AIC_{1}}$ (same for BIC) in terms of the strength of evidence against model 2 \citep{Kass1995,Tan2012}. The rule of thumb is that if $\Delta\leq2$, the favor degree is preferred; if $2<\Delta\leq5$, it is adjudged to be slight; and if $\Delta\geq5$, it is excluded.

For the AIC criterion, the utilization of different BAO datasets in combination with QSO yields varying outcomes. It is crucial to keep in mind that model selection provides quantitative information regarding the strength of evidence (or the level of support) rather than solely selecting a single model. The combination of 2D-BAO and 3D-BAO yields significantly disparate results, which merits particular attention. To show our results more clearly, we present a ranking of models based on AIC to determine the preference of observational data for these models in Table II. In the frame work of QSO+2D-BAO dataset, the favor degree based on AIC difference for GCG and  Case II models are preferred. For GEDE and PEDE models, they are adjudged to be slight. For $\Lambda$CDM, DGP, $\alpha$DE, NGCG and Case I models, their adjudged results are excluded. In the frame work of QSO+3D-BAO dataset, The DGP, $\alpha$DE, GCG and Case II models are adjudged to be preferred. For GEDE, NGCG and Case I models, they are slightly preferred by observational dataset. The $\Lambda$CDM and PEDE are adjudged to be excluded.
Combine the above description, we summarize that the two models, GCG and CASE II  models, performed best, and both are favored by the observational data. However, regardless of which dataset QSO+2D-BAO and QSO+3D-BAO are considered, AIC excludes the $\Lambda$CDM model. This is reasonable. Because AIC encourages good data fitting. Compared with $\Lambda$CDM model and others (two free parameters), GCG and Case II models (three free parameters) increase the number of free parameters and improve the goodness of fitting, but may try to avoid overfitting in four free parameters models.

The penalty term of BIC is larger than that of AIC, and when the number of samples is too large, it can effectively prevent the model complexity caused by too high precision.
When considering the BIC, similar to the AIC criterion, different results are obtained with varying BAO datasets. Specifically, QSO+2D-BAO dataset favors the PEDE and $\Lambda$CDM models while QSO+3D-BAO data supports the DGP model. Meanwhile, QSO+2D-BAO dataset slightly favors GEDE, GCG and Case II models, and  QSO+3D-BAO dataset   slightly favors $\Lambda$CDM and Case II models. There are four models to be excluded named DGP, $\alpha$DE, NGCG
and Case I model under the QSO+2D-BAO dataset. However, for QSO+3D-BAO dataset, there are six models to be excluded
$\alpha$DE, PEDE, GEDE, GCG, NGCG and Case I models. Combining the results of BIC under the two datasets,  we summarize that the $\Lambda$CDM and Case II models performs well.
It is obvious that models with more free parameters are less favored by current quasar and BAO observations. The evidences against for NGCG, Case I, and $\alpha$DE models are particularly notable for all data sets, indicating that the these three models are severely penalized by the BIC.

\section{Conclusion}
In this paper, we have evaluated the power of measurements of quasars covering sufficiently wide redshift range, on Brane-world models, Chaplygin Gas models, Emergent Dark Energy models and models with torsion, under the assumption of the spatial flatness of the Universe. The dynamics of these models are shown in Fig. 10. Considering that the dynamics of the above models are very similar at low redshift, the difference is significant at high redshift. Therefore, the newest large sample of QSO X-ray and UV flux measurements \citep{Lusso2020} were used as standard candles and provided a good opportunity to test models at
the ``redshift desert'' ($0.009<z<7.5$) that is not yet widely available through other observations. Meanwhile, with the aim to tighten the constraint from the combined QSOs datasets
and test the consistency with other observations, QSO+2D-BAO and QSO+3D-BAO datasets were also taken into account in this work. Here we summarize our main conclusions in more detail:

\begin{itemize}
\item For all non-standard cosmological models above, the QSO data alone are not able to provide effective constraints on model parameters, which is mainly related to the large dispersion of the $L_{X}-L_{UV}$ relation obtained from the 2421 QSO X-ray and UV flux measurements. With the addition of 2D and 3D-BAO datasets, we expect to be able to help break down the degeneracy between cosmological model parameters." The results were as expected, and the addition of BAO data helped us achieve this, a conclusion that can be clearly reflected in the $\alpha$DE, GCG and  GEDE models.

\item For all of the cosmological models considered, the constrained parameter precision using QSO+BAO data is significantly higher than that using BAO data alone. And the value of matter density parameter $\Omega_{m}$ implied by the QSO+BAO data is noticeably larger than BAO data, which is likely caused by the discrepancy between the QSO X-Ray and UV flux data and the $\Omega_{m}$ flat $\Lambda$CDM \citep{Risaliti2019}.

\item Our results show that the value of the Hubble constant varies considerably by considering different cosmological models, which means that the underlying model describing our real Universe remains unknown. From a theoretical point of view, alternative cosmological models may be an important way to alleviate or even solve the Hubble tension. Our work also supports that the Phenomenologically Emergent Dark Energy model and cosmological torsion models may alleviate the Hubble tension, the reported value of the Hubble constant lies between Planck's 2018 CMB observations and local measurements from the SH0ES collaboration obtained from QSO+BAO datasets combination, while other cosmological models all support that the Hubble constant tends to be closer to recent Planck 2018 results.

\item Based on the information criterion criterion, we find that $\Lambda$CDM model is not the most observed favored model. According to the AIC criterion, the utilization of different BAO datasets in combination with QSO yields varying outcomes, wherein the QSO+2D-BAO data favors the GCG model and the QSO+3D-BAO data favors the $\alpha$DE model.  For the BIC criterion, QSO+2D-BAO data favors the PEDE and $\Lambda$CDM models while QSO+3D-BAO data supports the DGP model. The evidence against NGCG and Case I models are particularly notable for all data sets, indicating that the NGCG model and  Case I models are severely penalized by the AIC and BIC. In addition, we find that, non-standard models with more free parameters are less favored by the available observations, which is the most unambiguous result of the current dataset.

\item It is necessary to note that,  the constrained results on $H_0$ for some considered models does not lie between the Planck 2018 results and local measurements from the SH0ES collaboration. For instance, in the frame work of  Case I model, all constrained results on $H_0$ in a very low value, much lower than Planck  reported. Based on information criterion, all criteria are unfriendly to this model. We conclude that the current datasets exclude this model. Therefore, based on the information criterion, it is beneficial for us to select or exclude some models to obtain the reliable results on $H_0$. This also applies to the GEDE model. However, there is a slight difference. Only BIC penalizes the GEDE model obtained using QSO+3D-BAO dataset and excludes this model. As we pointed above, this model has more free parameters, which causes BIC to penalize this model more, and there is a strong degeneracy between the $H_0$ and other model parameters, which affects the resulting $H_0$ constraint. In conclusion, we find that for models that are not excluded by the information criterion, the constraining results of $H_0$ are between the Planck 2018 results and local measurements from the SH0ES collaboration, which means that the information criterion can effectively help us to obtain more reliable constraining results and select cosmological models.

\end{itemize}

As a promising cosmological probe, quasars will play an increasingly important role in future model testing and have great potential to alleviate the tension problem of the Hubble constant at the high redshift. From the observational point of view, current COSMOS survey, XMM-Newton Serendipitous Source Catalog Data Release and future the Sloan Digital Sky Survey (SDSS) and XMM spectra  will bring us hundreds of thousands of quasars in the most optimistic discovery scenario, which will yield more high precision observational data with small uncertainties and dispersion in the future. On the other hand, We also expect to find more reasonable and accurate alternative models of the Universe in the future.

\acknowledgments
The authors are grateful the referee for careful reading the manuscript and numerous insightful comments.
Liu T.-H  was supported by National Natural Science Foundation of China under Grant No. 12203009; Chutian Scholars Program in Hubei Province (X2023007); Hubei Province Foreign Expert Project (2023DJC040).  Zheng X.-G. was supported by the National Natural Science Foundation of China under Grant No. 12103036;  Chutian Scholars Program in Hubei Province.


\label{lastpage}


%

\end{document}